\documentclass[12pt,epsf]{article}

\usepackage{amsfonts,amssymb,amsmath}
\usepackage{graphicx}
\usepackage{latexsym}
\usepackage{bm}
\usepackage{cite}
\usepackage{color}

\newcommand{\be}{\begin{equation}}
\newcommand{\ee}{\end{equation}}
\newcommand{\ba}{\begin{eqnarray}}
\newcommand{\ea}{\end{eqnarray}}

\begin{document}

\title{\textbf{Fubini instantons in curved space}}
\author{\small{\textsc{Bum-Hoon Lee}$^{a,b}$\footnote{\texttt{bhl@sogang.ac.kr}},\;\;
  \textsc{Wonwoo Lee}$^{b}$\footnote{\texttt{warrior@sogang.ac.kr}},\;\;
  \textsc{Changheon Oh}$^{c}$\footnote{\texttt{choh0423@gmail.com}},\;\;
  \textsc{Daeho Ro}$^{a}$\footnote{\texttt{dhro@sogang.ac.kr}},}\\ \small{and
  \textsc{Dong-han Yeom}$^{b,d}$\footnote{\texttt{innocent.yeom@gmail.com}}}\\
  \textit{\small{$^{a}$Department of Physics and BK21 Division, Sogang University, Seoul 121-742, Korea}}\\
  \textit{\small{$^{b}$Center for Quantum Spacetime, Sogang University, Seoul 121-742, Korea}}\\
  \textit{\small{$^{c}$Meerecompany, Gyeonggi-do 445-938, Korea}}\\
  \textit{\small{$^{d}$Yukawa Institute for Theoretical Physics, Kyoto University, Kyoto 606-8502, Japan}}}
\maketitle

\begin{abstract}
We study Fubini instantons of a self-gravitating scalar field. The
Fubini instanton describes the decay of a vacuum state under
tunneling instead of rolling in the presence of a tachyonic
potential. The tunneling occurs from the maximum of the potential,
which is a vacuum state, to any arbitrary state, belonging to the
tunneling without any barrier. We consider two different types of
the tachyonic potential. One has only a quartic term. The other has
both the quartic and quadratic terms. We show that, there exist
several kinds of new O(4)-symmetric Fubini instanton solution, which
are possible only if gravity is taken into account. One type of them has the structure with $Z_2$ symmetry. This type of the solution is possible only in the de Sitter background. We discuss on the interpretation of the solutions with $Z_2$ symmetry.
\end{abstract}
{\vspace{28pt plus 10pt minus 18pt}
 \noindent{\small\rm PACS numbers: 04.62.+v, 98.80.Cq\par}}

\newpage
\section{Introduction \label{sec1}}

The very first picture of an inflationary multiverse scenario was
proposed in Ref.\ \cite{alinde00}, in which it would seem that the
author wanted to suggest a universe without the cosmological
singularity problem using an interesting feature of
self-reproducting or regenerating exponential expansion of the
universe. A major development in this scenario was triggered by the
discovery of the eternal inflationary scenario \cite{eterinf, lin03,
guth00, wini09} and a paradigm for string theory landscape
\cite{land00, land01}. The eternal inflation is related to the
expanding false vacuum solution with a positive cosmological
constant, which in turn means that the inflation is eternal into the
future. If the theory has multiple minima then the false vacuum
state decays into the true vacuum state, i.e. the phase transition
proceeded {\it via} the nucleation of a vacuum bubble. In this scenario the universe is situated within some bubble called a pocket universe \cite{guth00} having a certain value of the cosmological constant and the whole universes are referred to as multiverse. The
description of self-reproduction including tunneling process and
random walk was combined into a scenario called recycling universe
\cite{gavil00}. These scenarios seem to provide an escape from the
question of the initial conditions of the universe, i.e. it seems to
be eternal into the past. Unfortunately, inflationary spacetimes
cannot be made complete in the past direction \cite{bvg}, even
though the universe is eternal into the future. There are still  interesting arguments on the beginning of the universe \cite{mvss}. The string theory landscape is a setting that involves a huge number of different metastable and stable vacua \cite{kklt, asdo}, originated from different choices of Calabi-Yau manifolds and generalized magnetic fluxes. The huge number of different vacua can be approximated by the potential of a scalar field. The important thing is the fact that, once the de Sitter vacuum can exist, the inflationary expansion is eternal into the future and has the possibility of self-reproduction.

On the other hand, there are theories of gauged $d=4$, $N=8$
supergravity having de Sitter(dS) solution, in which all SUSYs are
spontaneously broken. It is well known that the dS solution
corresponds to a M/sting theory solution with a non-compact $7$- or
$6$-dimensional internal space, in which a small value of the
cosmological constant stems from the $4$-form flux. The simplest
representative of these kind of theories has a tachyonic potential
with the dS maxima \cite{hull, klps, kl90}. The potential in the
vicinity of the maximum reduces to a form having a quadratic term,
that is not metastable but unstable. However, according to some
authors, the time for collapse giving rise to the tachyonic
potential can be much greater than the age of the universe for
anthropic reasoning. If the curvature radius of the potential in the
vicinity of the maximum is greater than that used in the above
theory, then that will be all together different story. The
supergravity analogue of the tachyonic potential could be
constructed also by using an exact supergravity solution
representing the $D_p$-$\bar{D}_p$ system \cite{hskim}.

From the above scenarios, the study of the possibility of the
tunneling process for the potential with stable and metastable
vacua, or even tachyonic behavior has acquired renewed interest. In
the present paper, we will study the tunneling process under a
simple tachyonic potential governed by a quartic term both without
the quadratic term and with the term as a toy model. To obtain the
general solution including the effect of the backreaction, we solve
the coupled equations for the gravity and the scalar field
simultaneously. Although the model has a tachyonic potential, it
might still be an useful example to show how the tunneling process
occurs in various shapes of the potential provided by the above
scenarios.

A quantum particle can tunnel through a finite potential barrier
{\it via} the so-called barrier penetration. This process can be
described by the Euclidean solution obeying
appropriate boundary conditions. There exist two kinds of Euclidean
solutions describing quantum tunneling phenomena. One corresponds to
an instanton solution representing a stable pseudoparticle
configuration characterized by the existence of a nontrivial
topological charge. It does not change even if we continuously
deform the field, as long as the boundary conditions remain the
same. The instanton solution corresponds to the minimum of the
Euclidean action to pass from the initial to final state \cite{bpst01}. The solution, in case of a double well potential, describes a general shift in the ground state energy of the classical vacuum due to the presence of an additional potential well, then lifting the so-called classical degeneracy. The other is a bounce solution representing an unstable nontopological configuration that corresponds to a saddle point rather than a minimum of the Euclidean action. The second derivative of the Euclidean action around the bounce has one negative eigenvalue which leads to the imaginary part of the energy. The existence of the negative eigenvalue implies that the vacuum state is unstable, i.e. the state decays into other states \cite{colm02}.

The Euclidean solutions can also mediate phase transitions. The
phase transition describes the sudden change of a physical system
from one state to another. The transition are of two different types
transition accompanied by temperature or zero temperature. The
competition between the entropy and the energy terms in the
thermodynamic potential cause thermal phase transitions in which
dynamics is irrelevant. In the modern classification scheme, thermal
phase transitions are divided into two broad categories either with
a discontinuous jump in the first-order derivatives of the free
energy or without it. A first-order phase transition is
characterized by the discontinuity in the first derivative of the
free energy and is associated with the existence of latent heat,
whereas a $n$th-order phase transition is characterized by the
continuity in the first derivative while there is a discontinuity in
the $n$th-order derivative. A quantum phase transition describes a
transition between different phases by quantum fluctuation, which
occurs at zero temperature, unlike the case of a thermal phase
transition which is governed by a thermal fluctuation \cite{herbut}.

To simplify things, we consider an asymmetric double well potential
to distinguish two different phase transitions at zero temperature.
If the initial state is the metastable vacuum state and the
tunneling occurs from that state to the other vacuum state, then the
transition corresponds to a tunneling process \cite{voloshin,
col002, bnu02, par02, klee07, ll090}. On the other hand, if the
initial state is the local maxima of the potential and the field is
rolling down to one vacuum state continuously rather than any
discontinuous jump, then the transition corresponds to the rolling.
However, one more channel exists as tunneling and that corresponds
to the one without a barrier. In this kind of transition, the
initial state on the top of a potential can tunnel to the other
state rather than rolling down the potential \cite{fubi01, linde00,
klee08, lllo2}. There are two different kind of transitions in this
case. One is the tunneling without a barrier representing the
tunneling from the local maximum of the potential to the vacuum
state \cite{klee08, ljps, lllo2, kss00}. Recently, an analytic study
on this type of solution was performed in \cite{kss00}. The other is
a tunneling without a barrier representing the tunneling from the
maximum of the potential to any arbitrary state. This case
corresponds to the Fubini instanton \cite{fubi01, linde00}, where
the tachyonic potential is employed. Can we describe the rolling
corresponding to the transition between the initial metastable
vacuum state and the other final vacuum state? This may look similar
to a superfluid motion by the liquid helium. Although to establish
the phase transition corresponding to the superfluid motion is
itself a very challenging problem, we concentrate on the Fubini
instantons in this work.

The Fubini instanton \cite{fubi01, linde00} describes the decay of a vacuum state by the quantum phase transition instead of rolling down the tachyonic potential consisted of a quartic term only. On the other hand, one can consider a tachyonic potential consisted of a quadratic term only, the point $\Phi=0$ is unstable. A small perturbation will cause it to roll down the hill of the potential. Originally, it was Fubini who introduced a fundamental scale of hadron phenomena by means of the dilatation noninvariant vacuum state in the framework of a scale invariant Lagrangian field theory \cite{fubi01}. However, the solution is a one-parameter family of instanton solutions representing a tunneling without a barrier as an interpolating solution from the maximum of the potential to any
arbitrary state. The instanton solution was studied in a conformally
invariant model, i.e. a fixed background was used and the effect of
the backreaction by instantons was neglected \cite{gmst, khleb,
loran01}. This is a good approximation, when the variation of the
potential during the transition is much smaller than the maximum of
the potential. The instanton has gained much interest now-a-days in
the context of anti-de Sitter(AdS)/conformal field theory
correspondence \cite{hapet, bara}.

The paper is organized as follows: In Sec.\ 2, we review the Fubini
instanton in the absence of gravity. We present numerical solutions
including the Euclidean energy density as an example and analyze the
structure of the solution in the theory with a potential having only
the quartic self-interaction term. We stress the fact that there is
no such solutions with the potential containing both the quartic and
the quadratic terms. In Sec.\ 3, we show that the instanton
solutions exist in the curved space. We perform a numerical study to
solve the coupled equations for the gravity and the scalar field
simultaneously. We show that there exist numerical solutions without
oscillation in the initial AdS space in the potential with only the
quartic term. We also show that there exist numerical solutions in
the potential both with the quartic and the quadratic terms
irrespective of the value of the cosmological constant, which is
possible only when the gravity is switched on. In order to estimate
the decay rate of the background state, we compute the action
difference between that of the solution and the background obtained
by numerical means. We present an oscillating numerical
solutions in the potential with only the quartic term with various values of the cosmological constant. One type of these solutions has the structure with $Z_2$ symmetry. We will discuss on the interpretation of the solutions with $Z_2$ symmetry in the final Section. We analyze the behavior of the solutions using the phase diagram method. In Sec.\ 4, to observe the dynamics of the solutions, we briefly sketch the causal structures of the solutions in the Lorentzian spacetime. Finally in Sec.\ 5, we summarize and discuss our results.

\section{Fubini instanton in the absence of gravity  \label{sec2}}

\begin{figure}[t]
\begin{center}
\includegraphics[width=6.0in]{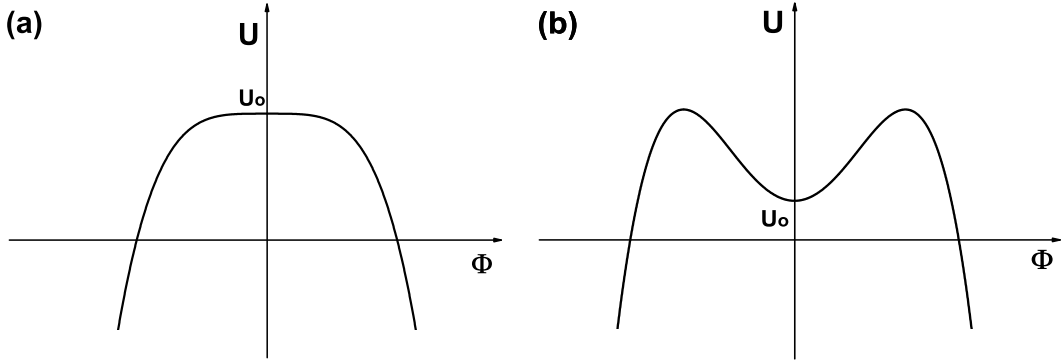}
\end{center}
\caption{\footnotesize{Potentials for the case of (a) Fubini
instantons and (b) generalized Fubini instantons.}}
\label{fig:fig01}
\end{figure}

One can consider the following action in the absence of gravity
\begin{equation}
S= \int_{\mathcal M} \sqrt{-g} d^4 x \left[
-\frac{1}{2}{\nabla^\alpha}\Phi {\nabla_\alpha}\Phi -U(\Phi)\right] , \label{min-action}
\end{equation}
where $g=\det\eta_{\mu\nu}$, $\eta_{\mu\nu}=diag(-1,1,1,1)$ is the Minkowski metric, and the tachyonic potential has a quartic self-interaction term and also a quadratic term as follows:
\begin{equation}
U(\Phi)=-\frac{\lambda}{4} \Phi^4 + \frac{m^2}{2}\Phi^2  +U_o , \label{min-po}
\end{equation}
where $m^2 > 0$ and $\lambda > 0$. The plots of potentials (a)
without the quadratic term and (b) with the quadratic term are shown
in Fig.\ \ref{fig:fig01}. The potential has a metastable vacuum
state at $\Phi=0$ and no other stationary state in Fig.\
\ref{fig:fig01}(a), while Fig.\ \ref{fig:fig01}(b) illustrates that
the potential has a local minimum at $\Phi=0$ and two maxima
\cite{linde00, kt10, yy9}. In both the cases, the potential is not
bounded from below.

Before going to the tunneling problem in four dimensions, we briefly
describe the problem in one dimension. One can consider the simplest
quantum tunneling problem in one dimension. Quantum field theory in
one dimension is nothing but ordinary quantum mechanics. In case of
$m^2 = 0$, the amplitude for transmission obeys the WKB formula in
the semiclassical approximation, in which
$\Phi_{\pm}={\pm}(\frac{4U_o}{\lambda})^{1/4}$ are the classical
turning points. On the other hand, the double-hump potential with
$m^2 > 0$ and $U_o =0$ can be considered as an inverted double-well
potential for a bounce solution representing the tunneling from
$\Phi=0$ to $\Phi_{\pm}= \pm m\sqrt{{2}/{\lambda}}$. The solution is
given by $\Phi_s(\tau)= \pm m\sqrt{{2}/{\lambda}} \mathrm{sech}
[m(\tau-\tau_o)]$, where $\tau_o$ is an integration constant. The
bounce solutions can be easily understood in the Euclidean space. The particle can only reach the point $\Phi=0$ at $\tau= \pm \infty$ and
it bounces off  $\Phi_{\pm}= \pm m\sqrt{{2}/{\lambda}}$ at $\tau=0$
with a vanishing velocity \cite{bewulik}.

We now turn to the tunneling problem in four dimensions. It is an
well-known fact that the massless theory has an instanton
\cite{fubi01}. Actually, the instanton corresponds to the bounce
solution representing the decay of the background vacuum state. The
equation of motion with $O(4)$ symmetry, obtained by varying the
Euclidean action, is then;
\begin{equation}
\frac{d^2\Phi}{d\eta^2} + \frac{3}{\eta} \frac{d\Phi}{d\eta}
=-\frac{d(-U)}{d\Phi}\,, \label{min-eeq}
\end{equation}
where $\eta(=\sqrt{\tau^2 + x^2})$ plays the role of the evolution parameter in Euclidean space and the second term in the left-hand side plays the role of a damping term. The boundary conditions are
\begin{equation}
\frac{d\Phi}{d\eta}\Big|_{\eta=0}=0 \,\,\,\, {\rm and}\,\,\,\,
\Phi|_{\eta=\infty}=0\,.
\label{ourbc-2}
\end{equation}
The particle in the classical mechanics problem starts at
$\Phi=\Phi_o$ with zero velocity in the inverted potential, and
stops at $\Phi|_{\eta=\infty}=0$ without any oscillation.

For the potential with $m^2 = 0$, the analytic solution of the
Fubini instanton has the form
\begin{equation}
\Phi(\eta)= \sqrt{\frac{8}{\lambda}}\frac{b}{\eta^2+b^2}\,,
\label{min-asol}
\end{equation}
where $\eta$ is the radial length in the Euclidean space, $b$ is any
arbitrary length scale which characterizes the size of the instanton
and is related to the initial value $\Phi_o$. In addition, the value
of the scalar field of the center of the solution depends on $b$ as
$\Phi(0)=\sqrt{\frac{8}{\lambda}}\frac{1}{b}$. This solution was used in the related perturbation theory \cite{lipatov}.

\begin{figure}[t]
\begin{center}
\includegraphics[width=3.0in]{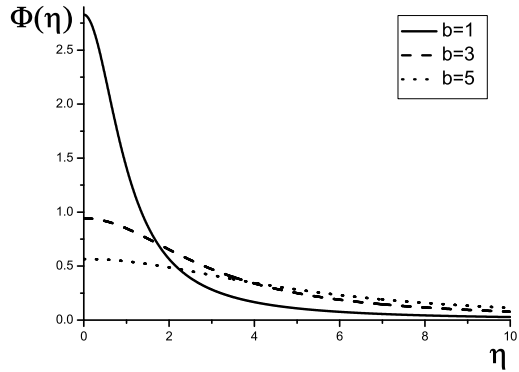}
\end{center}
\caption{\footnotesize{The analytic solution of the Fubini instanton
in absence of gravity.}} \label{fig:fig02}
\end{figure}

The characteristic behavior of the analytic solution is ploted in
Fig.\ \ref{fig:fig02} in terms of the value of the parameter $b$. We
take $\lambda=1$ for all the cases. The solid line denotes the
solution with $b=1$, the dashed line with $b=3$, and the dotted line
with $b=5$. The corresponding Euclidean action is given by
\begin{equation}
S_{\mathrm{E}}= \frac{32\pi^2 b^2}{\lambda} \int^{\infty}_{0}
\frac{\eta^5(1-\frac{b^2}{\eta^2})}{(\eta^2+b^2)^4} d\eta =
\frac{8\pi^2}{3\lambda}\,, \label{min-act}
\end{equation}
where the action does not depend on the parameter $b$ due to the
consequence of the conformal invariance of the potential and we take
that value to be $U_o=0$. The action has the same value irrespective
of the starting point $\Phi_o$. In other words, the tunneling from
the maximum of the potential to any arbitrary state always happens
with same probability.

\begin{figure}[t]
\begin{center}
\includegraphics[width=6.0in]{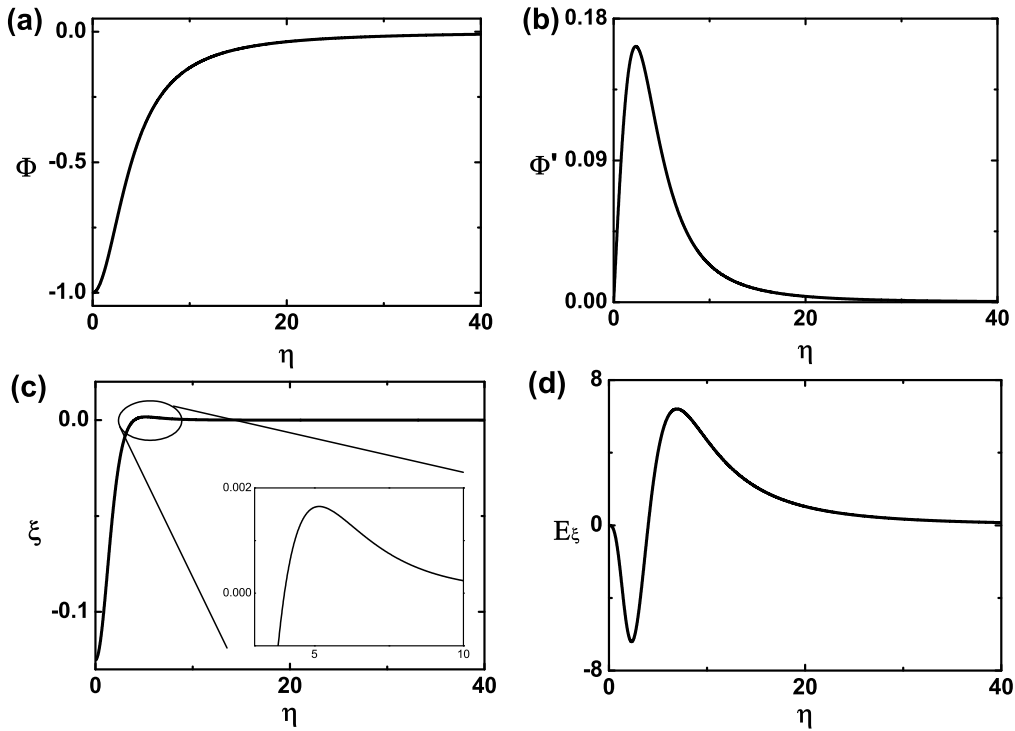}
\end{center}
\caption{\footnotesize{(a) The numerical solution for $\Phi$ in the
case of $m^2=0$, (b) the variation of $\Phi'$ with respect to
$\eta$, (c) the energy density $\xi$ and (d) the Euclidean energy
$E_{\xi}$ evaluated at constant $\eta$.}} \label{fig:fig03}
\end{figure}

The numerical solutions for $\Phi$ and $\Phi'$ and the Euclidean
energy including the density variation with $\eta$ are as shown in
Fig.\ \ref{fig:fig03}. Figure \ref{fig:fig03}(a) illustrates the
numerical solution for $\Phi$, in which the initial value set as
$\Phi_o =-1$ and the solution asymptotically approaches the value
$\Phi = 0$. Figure \ref{fig:fig03}(b) illustrates $\Phi'$ with
respect to $\eta$. There is a peak of $\Phi'$ near $\eta = 2.31$.
Figure \ref{fig:fig03}(c) depicts the volume energy density, when
the density has got the form $\xi=[\frac{1}{2}\Phi'^2+U]$. The lower
right box in the same figure shows the magnification of the small
region clearly representing the existence of a smooth hill. The
smooth peak of the volume energy density exists at $\eta=5.17$.
There is a disagreement between the location of the peak for the
energy density and that for $\Phi'$. It clearly reveals the fact
that the position with the maximum value for $\Phi'$ is still not
the same as the maximum of the energy density due to non-trivial
contribution coming from the potential, $U=-\frac{\lambda}{4}\Phi^4
$. Figure \ref{fig:fig03}(d) shows the Euclidean energy $E_{\xi}$
for each slice of constant $\eta$. The Euclidean energy signifies
the value of energy after the full integration of variables except
for $\eta$ in the present work, $E_{\xi}=2\pi^2\eta^3\xi$. There are
one minimum and one maximum point for $E_{\xi}$. The location of the
minimum of $E_{\xi}$ is around $\eta=2.31$, whereas that of the
maximum is near $\eta=6.93$. Ironically, the location of the minimum
of $E_{\xi}$ coincides with that of the maximum of $\Phi'$. These
solutions can be considered as a ball consisting of only a thick
wall except for one point at the center of the solution with a lower
arbitrary state than the outer vacuum state unlike a vacuum bubble
that consists of an inside part with a lower vacuum state and a
wall.

For a theory with $m^2 > 0$, the conformal invariance is broken and
any solution with a finite action is forbidden by scaling
argument. In other worlds, the particle can not have enough energy
to reach the hill overcome the barrier near $\Phi=0$ since the
damping term has got a large value due to a large value of $\Phi'$
near the initial point \cite{affle01}.

\section{Fubini instantons of a self-gravitating scalar field \label{sec3}}

Let us consider the following action:
\begin{equation}
S= \int_{\mathcal M} \sqrt{-g} d^4 x \left[ \frac{R}{2\kappa}
-\frac{1}{2}{\nabla^\alpha}\Phi {\nabla_\alpha}\Phi -U(\Phi)\right] +
\oint_{\partial \mathcal M} \sqrt{-h} d^3 x \frac{K-K_o}{\kappa}\,,
\label{f-action}
\end{equation}
where $g=\det g_{\mu\nu}$, $\kappa \equiv 8\pi G$, $R$ denotes the
scalar curvature of the spacetime $\mathcal M$, $h$ is the induced
boundary metric, $K$ and $K_o$ are the traces of the extrinsic
curvatures of $\partial \mathcal M$ for the metric $g_{\mu\nu}$ and
$\eta_{\mu\nu}$, respectively. The second term on the right-hand
side is the boundary term \cite{York}. It is necessary to have a
well-posed variational problem including the Einstein-Hilbert term.
Here we adopt the notations and sign conventions of Misner, Thorne
and Wheeler \cite{misner}.

We study the creation process of Fubini instantons in curved
spacetime. In the first place, we consider the massless case, and
then we will also consider generalized Fubini instantons, the
so-called massive case (see the form of the potential in Eq.\
(\ref{min-po})). The cosmological constant is given by $\Lambda =
\kappa U_{0}$, such that background space will be dS, flat or AdS
depending on the signs of $U_{0}$.

In order to solve the coupled equations, we assume an $O(4)$
symmetry for the geometry and the scalar field similar to Ref.\
\cite{bnu02}
\begin{equation}
ds^{2} = d\eta^{2} + \rho^{2}(\eta) \left[ d\chi^{2} + \sin^{2}\chi \left(
d\theta^{2} + \sin^{2}\theta d\phi^{2} \right) \right] \,.
\end{equation}
And then, $\Phi$ and $\rho$ depends only on $\eta$, and the
Euclidean equation can be written respectively as follows:
\begin{equation}
\Phi'' + \frac{3\rho'}{\rho}\Phi'=\frac{dU}{d\Phi} \,\,\, {\rm and} \,\,\,
\rho'' = - \frac{\kappa}{3}\rho (\Phi'^2 +U)\,, \label{erho}
\end{equation}
and the Hamiltonian constraint is then given by
\begin{equation}
\rho'^2 - 1 - \frac{\kappa\rho^2}{3}\left(\frac{1}{2}\Phi'^{2}-U\right) = 0 \,.
\label{eqcon}
\end{equation}
In order to yield a meaningful solution, the constraint requires a
delicate balance among all the different terms. Otherwise the
solution can yield qualitatively incorrect behavior \cite{berg}.

To solve the Eqs.\ (\ref{erho}), we have to impose suitable boundary
conditions. When the gravity is switched off, boundary conditions
for the Fubini instanton are $\frac{d\Phi}{d\eta}\Big|_{\eta=0}=0$
and $\Phi|_{\eta=\infty}=0$ as in Ref.\ \cite{fubi01}. While gravity
is taken into account, we can write boundary conditions as follows:
\begin{equation}
\rho|_{\eta=0}=0, \,\,\,\, \frac{d\rho}{d\eta}\Big|_{\eta=0}=1, \,\,\,\,
\frac{d\Phi}{d\eta}\Big|_{\eta=0}=0, \,\,\,\, {\rm and}\,\,\,\,
\Phi|_{\eta=\eta_{max}}=0 \,,\label{ourbc-2}
\end{equation}
where $\eta_{max}$ is the maximum value of $\eta$. For the flat and AdS background $\eta_{max}=\infty$, while $\eta_{max}$ is finite for the dS background. The first condition is to obtain a geodesically complete spacetime. The second condition is nothing but Eq.\ (\ref{eqcon}). The third condition is the regularity condition as can be seen from the first equation in Eq.\ (\ref{erho}). One should find the undetermined initial value of $\Phi$, i.e. $\Phi|_{\eta=0}=\Phi_o$, using the undershoot-overshoot procedure \cite{col002, lllo2}, to satisfy the fourth condition $\Phi|_{\eta=\eta_{max}}=0$. We employ these conditions for Fubini instantons in Sec.\ III A, B.

If the background space is dS, we can impose conditions specified at $\eta = 0$ and $\eta=\eta_{max}$. For this purpose, we choose the values of the field $\rho$ and derivatives of the field $\Phi$ as follows:
\begin{equation}
\rho|_{\eta=0}=0 ,
\,\,\,\, \rho|_{\eta =\eta_{max}} =0, \,\,\,\,
\frac{d\Phi}{d\eta}\Big|_{\eta=0}=0, \,\,\,\, {\rm and}\,\,\,\,
\frac{d\Phi}{d\eta}\Big|_{\eta =\eta_{max}} = 0. \label{ourbc-3}
\end{equation}
The first two conditions are for the background space. The last two conditions are for the scalar field. In general, the solutions satisfying Eq.\ (\ref{ourbc-3}) do not guaranty $\Phi_{\eta=\eta_{max}}$ to be zero. For the solution having $\Phi_{\eta=\eta_{max}}=0$, the conditions Eq.\ (\ref{ourbc-3}) are equivalent to the conditions Eq.\ (\ref{ourbc-2}). If $\Phi_{\eta=\eta_{max}} = \pm \Phi_o$, they represent completely new type of solutions with $Z_2$ symmetry. We will discuss this case more in detail in Sec.\ III C.

In order to solve the Euclidean field Eqs.\ (\ref{erho}) and
(\ref{eqcon}) numerically, we rewrite the equations in terms of
dimensionless variables as in Ref.\ \cite{lllo2}. In the present
work, we employ the shooting method using the adaptive step size
Runge-Kutta as the integrator similar to the treatment in
Ref.\ \cite{NR}. For this procedure we choose the initial values of
$\tilde{\Phi}(\tilde{\eta}_{\mathrm{initial}})$,
$\tilde{\Phi}'(\tilde{\eta}_{\mathrm{initial}})$,
$\tilde{\rho}(\tilde{\eta}_{\mathrm{initial}})$, and
$\tilde{\rho}'(\tilde{\eta}_{\mathrm{initial}})$ at
$\tilde{\eta}=\tilde{\eta}_{\mathrm{initial}}$ as follows:
\begin{eqnarray}
\tilde{\Phi}(\tilde{\eta}_{\mathrm{initial}}) &\sim& \tilde{\Phi}_{o} - \frac{\epsilon^2}{8}\tilde\Phi_o(\tilde\Phi^2_o-1) +  \cdot\cdot\cdot \,,  \nonumber \\
\tilde{\Phi}'(\tilde{\eta}_{\mathrm{initial}}) &\sim& - \frac{\epsilon}{4}\tilde\Phi_o(\tilde\Phi^2_o-1) + \cdot\cdot\cdot \,,  \label{nbcon2} \\
\tilde{\rho}(\tilde{\eta}_{\mathrm{initial}}) &\sim& \epsilon + \cdot\cdot\cdot \,,  \nonumber \\
\tilde{\rho}'(\tilde{\eta}_{\mathrm{initial}}) &\sim& 1 + \cdot\cdot\cdot \,, \nonumber
\end{eqnarray}
where $\tilde{\eta}_{\mathrm{initial}} = 0+\epsilon$ for $\epsilon
\ll 1$. The minus sign in front of the second formula is due to the
negative value of the $\tilde\Phi''$ determined by the sign of
$dU/d\Phi$ at $\tilde\eta=0$. However, the initial value of
$\tilde{\Phi}'$ is taking to be positive in the present work. Once
we specify the initial value $\tilde{\Phi}_{0}$, the remaining
conditions can be exactly determined from Eqs.\ (\ref{nbcon2}). Furthermore we impose additional conditions implicitly. To avoid a singular solution at $\tilde{\eta} = \tilde{\eta}_{\mathrm{max}}$ for the Euclidean field equations and to demand a $Z_2$ symmetry, the conditions
$d\tilde{\Phi}/{d\tilde{\eta}} \rightarrow 0$ and
$\tilde{\rho}\rightarrow 0$ as $\tilde{\eta}\rightarrow
\tilde{\eta}_{\mathrm{max}}$ are needed in the next section. In this
work, we require that the value of $d\tilde{\Phi}/{d\tilde{\eta}}$
goes to a value smaller than $10^{-6}$ as $\tilde{\eta} \rightarrow
\tilde{\eta}_{\mathrm{max}}$, as the exact value of
$\tilde{\eta}_{\mathrm{max}}$ is not known \cite{lllo2}.
The parameter $\tilde\kappa$ is the ratio between the gravitational
constant or Planck mass and the mass scale in the theory,
$\tilde{\kappa}=\frac{m^2}{\lambda}\kappa =\frac{8\pi
m^2}{M^{2}_{p}\lambda}$, and the parameter
$\tilde{\kappa}\tilde{U}_o$ is related to the rescaled cosmological
constant $\Lambda/m^2$.

To find the probability of the instanton solution, we only consider
the Euclidean action for the bulk part in Eq.\ (\ref{f-action}) to
get,
\begin{equation}
S_E= \int_{\mathcal M} \sqrt{g_{\mathrm{E}}} d^4 x_{\mathrm{E}}
\left[ -\frac{R_\mathrm{E}}{2\kappa} +\frac{1}{2}\Phi'^2 +U \right]
=  2\pi^2 \int \rho^3 d\eta [-U]\,, \label{euclac}
\end{equation}
where $R_{\mathrm{E}} =6[1/\rho^2 - \rho'^2/\rho^2 - \rho''/\rho]$.
We used Eqs.\ (\ref{erho}) and (\ref{eqcon}) to arrive at this. The
volume energy density has the form: $\xi=-U$, which has a different
sign compared to the sign of the density used in Ref.\ \cite{lllo2}.
The Euclidean energy signifies the energy value after the full
integration of variables except for $\eta$ in the present case as
$E_{\xi}=2\pi^2\rho^3\xi$.

In the beginning, we obtain the numerical solution for $m^2=0$. And
then we obtain the numerical solution for $m^2 > 0$. We call the
space dS when the initial vacuum state has a positive cosmological
constant, $U_o > 0$, flat when $U_o = 0$ and AdS when $U_o < 0$.

The rate of decay of a metastable state can be evaluated in terms of
the classical configuration and represented as $Ae^{-B}$ in this
approximation, in which the leading semiclassical exponent $B
=S^{\mathrm{cs}} - S^{\mathrm{bg}}$ is the difference between the
Euclidean action corresponding to the classical solution
$S^{\mathrm{cs}}$ and the background action $S^{\mathrm{bg}}$. The
prefactor $A$ is evaluated from the Gaussian integral over
fluctuations around the background classical solution \cite{ccol, coea}.

\begin{figure}[t]
\begin{center}
\includegraphics[width=6.in]{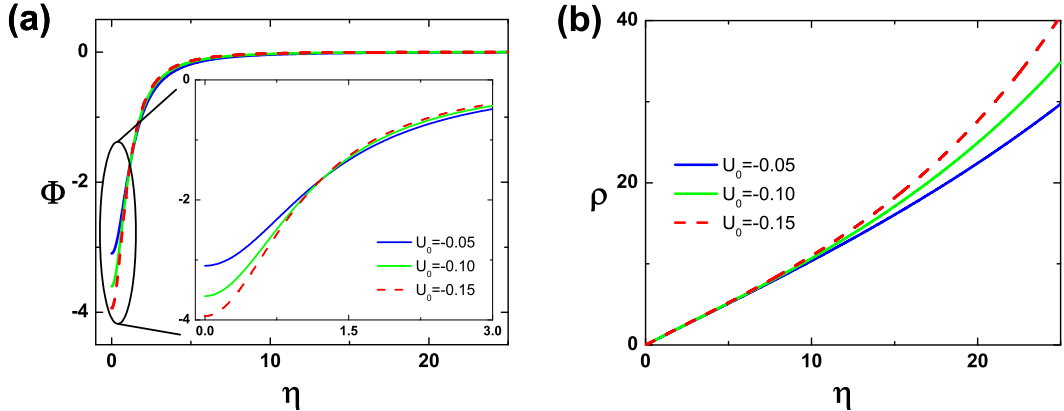}
\end{center}
\caption{\footnotesize{(color online). The numerical solutions of
Fubini instantons with $m^2=0$ in the AdS space.}} \label{fig:fig04}
\end{figure}

\begin{table}
\begin{center}
\renewcommand{\arraystretch}{1.5}
\begin{tabular}{cccccc}
  \noalign{\hrule height0.8pt}
  $\tilde{U}_{0}$ & $\tilde{\Phi}_{0}$ & Color of plot & $S^{\mathrm{cs}}$ & $S^{\mathrm{bg}}$ & B \\
  \hline
  $-0.05$ & $-3.09706$ & Red & $1.38470\times10^5$ & $1.36162\times10^5$ & $2.30801\times10^3$ \\
  $-0.10$ & $-3.60269$ & Green & $3.91419\times10^5$ & $3.82936\times10^5$ & $8.48298\times10^4$ \\
  $-0.15$ & $-3.93551$ & Blue & $8.25148\times10^5$ & $8.03225\times10^5$ & $2.19226\times10^4$ \\
  \noalign{\hrule height0.8pt}
\end{tabular}
\end{center}
\caption{\footnotesize{The dimensionless variables and color of plot
used and the actions obtained in Fig.\ \ref{fig:fig04}.}}
\label{table1}
\end{table}

\subsection{Solutions without oscillation}

We perform the numerical work with $m^2=0$ and take $\tilde{\kappa} =0.1$. The solutions without oscillation are only possible in the initial AdS background as shown in Fig.\ \ref{fig:fig04}. We guess that there is no solution without any oscillation for the initial flat and dS background. In the given $\tilde{\kappa}$, there may exist the phase space of solutions having the region of an arbitrary $\Phi_o$. If $\tilde{\kappa}$ is increased, the oscillating behavior is appearing in the phase space of solutions \cite{llry}.

Figure 4(a) illustrates the solution for $\tilde{\Phi}$, in which the right box in the same figure shows the magnification of a small region representing the initial values of $\tilde{\Phi}$ and the behavior of the curves. The curves move upwards with increasing value of $\tilde{U}_o$, then overlap near $\tilde{\eta}=1$, and more downwards with increasing value of $\tilde{U}_o$. Figure 4(b) shows the solutions of $\tilde{\rho}$. The curves move downwards with increasing $\tilde{U}_o$. The shape of the numerical solution $\tilde{\rho}$ can be easily understood if one thinks of the shape of the solution in a fixed AdS space as $\rho =
\sqrt{\frac{3}{\Lambda}}\sinh\sqrt{\frac{\Lambda}{3}} \eta$. Table 1
shows the dimensionless variables and the color of plot used, and
also the actions obtained from Fig.\ \ref{fig:fig04}. From the
numerical data, one can easily see that the magnitude of
$\tilde{\Phi}_o$ approaches the vacuum state $\tilde{\Phi}=0$ as
$\tilde{U}_o$ approaches a vanishing value. The vanishing of
$\tilde{U}_o$ means the background geometry which serves as the
initial vacuum state is flat. The action difference $\tilde{B}$ between the action of the solution $\tilde{S}^{\mathrm{cs}}$ and that of the background $\tilde{S}^{\mathrm{bg}}$ has positive values. We carry out the action integral in the range $0 \leqq \tilde{\eta} \leqq 25$ numerically as the action difference $\tilde{B}$ diverges to infinity if we perform the integration for an infinite $\tilde{\eta}$ value. This divergence is due to the fact that the size of the solution including the outside part in the evolution parameter space decrease compared to the size of the initial background similar to what happens for the case of the nucleation of a vacuum bubble. In the analytic computation, the outside part and the background are simply canceled at the same radius. In the present numerical work, it is difficult to decide the exact size of the solution. Thus we straightforwardly compute the action difference and then the difference $\tilde{B}$ has got an approximate behavior $\delta(\sinh^3{\tilde{\eta}})= 3\sinh^2\tilde{\eta} \cosh\tilde{\eta}$ which cause the divergence at infinity. If this minor error is cured, the action difference has a finite value.

\begin{figure}[t]
\begin{center}
\includegraphics[width=6.in]{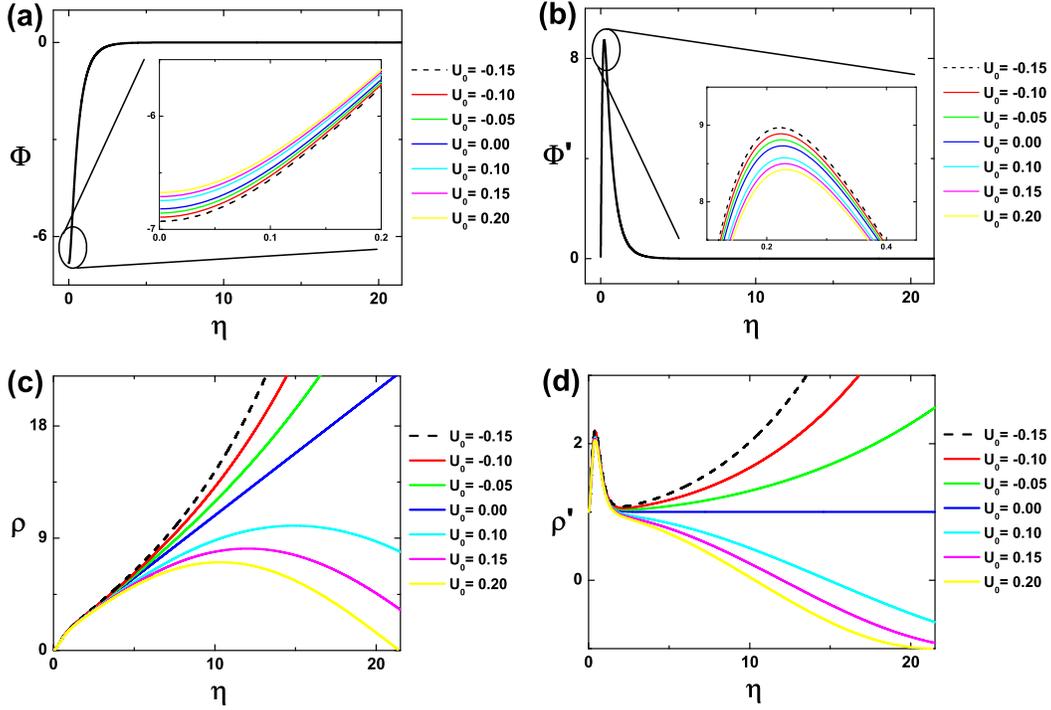}
\end{center}
\caption{\footnotesize{(color online). The numerical solution of
$\Phi$, the derivative of $\Phi$ with respect to $\eta$, $\rho$, and
the derivative of $\rho$ with respect to $\eta$ for the generalized
Fubini instantons with $m^2 > 0$.}} \label{fig:fig05}
\end{figure}

\begin{table}
\begin{center}
\renewcommand{\arraystretch}{1.5}
\begin{tabular}{cccccc}
  \noalign{\hrule height0.8pt}
  $\tilde{U}_{0}$ & $\tilde{\Phi}_{0}$ & Color of plot & $S^{\mathrm{cs}}$ & $S^{\mathrm{bg}}$ & B \\
  \hline
  $-0.15$ & $-6.92872$ & Black & $8.70060\times10^7$ & $4.13605\times10^7$ & $4.56455\times10^7$ \\
  $-0.10$ & $-6.89194$ & Red & $1.00592\times10^7$ & $7.78648\times10^6$ & $2.27277\times10^6$ \\
  $-0.05$ & $-6.85532$ & Green & $1.17726\times10^6$ & $9.75934\times10^5$ & $2.01329\times10^5$ \\
  $0.00$ & $-6.81885$ & Blue & $2.18938\times10^2$ & $0$ & $2.18938\times10^2$ \\
  $0.10$ & $-6.74631$ & Sky blue & $-2.61086\times10^4$ & $-2.63187\times10^4$ & $2.10070\times10^2$ \\
  $0.15$ & $-6.71021$ & Pink & $-1.73397\times10^4$ & $-1.75457\times10^4$ & $2.05942\times10^2$ \\
  $0.20$ & $-6.67421$ & Yellow & $-1.29572\times10^4$ & $-1.31591\times10^4$ & $2.01936\times10^2$ \\
  \noalign{\hrule height0.8pt}
\end{tabular}
\end{center}
\caption{\footnotesize{The dimensionless variables and color of plot
used and the actions obtained in Fig.\ \ref{fig:fig05}.}}
\label{table2}
\end{table}

Now we perform the numerical work with $m^2 > 0$ and take $\tilde{\kappa} =0.3$. This type of solutions belongs to usual tunneling with a barrier. We obtained the numerical solutions with an arbitrary cosmological constant as shown in Fig.\ \ref{fig:fig05}. The solutions are only possible for specific $\tilde{\Phi}_o$s.

The figures represent the vary fact that the solution is only possible in curved spacetime irrespective of the value of the cosmological constant. Figure 5(a) illustrates the solution of $\tilde{\Phi}$. The upper right box in the same figure shows the magnification of the small region representing behavior of the curves which move to the left with an increase in $\tilde{U}_o$. Figure \ref{fig:fig05}(b) shows $\tilde{\Phi}'$ with respect to $\tilde{\eta}$. The upper right box in the same figure shows the magnification of the small region representing behavior of the curves moving below with increasing value of $\tilde{U}_o$. Figure \ref{fig:fig05}(c) illustrates the solutions of $\tilde{\rho}$. The curves move downwards with increasing value of $\tilde{U}_o$. The shape of the numerical solutions of $\tilde{\rho}$ can be easily understood if one consider a fixed space. In the fixed flat space, $\rho = \eta$. In the dS
space, $\rho = \sqrt{\frac{3}{\Lambda}}\sin\sqrt{\frac{\Lambda}{3}}
\eta$. In the AdS space, $\rho =
\sqrt{\frac{3}{\Lambda}}\sinh\sqrt{\frac{\Lambda}{3}} \eta$.
Figure \ref{fig:fig05}(d) depicts the variation of $\tilde{\rho}$
with respect to $\tilde{\eta}$. The curves move below with increase in $\tilde{U}_o$. The horizontal line with $\tilde{U}_o=0$ indicate a flat space with $\tilde{\rho}'=1$. Table \ref{table2} shows the
dimensionless variables and color of the plot used among with the
action obtained from Fig.\ \ref{fig:fig05}. From the numerical data,
one can infer that the magnitude of $\tilde{\Phi}_o$ decreases as
$\tilde{U}_o$ increases. We carry out the action integral in the
range $0 \leqq \tilde{\eta} \leqq 30.58$ numerically. In the dS
space, the solution and the background have their own periods for
$\tilde{\eta}$, which we take the period as the integration limit. For the background dS space, we take $\tilde{\eta} = \pi
\sqrt{\frac{3}{\tilde{\kappa}\tilde{U}_o}}$. The action for
$\tilde{S}^{\mathrm{cs}}$ and $\tilde{S}^{\mathrm{bg}}$ are positive or zero as long as $\tilde{U}_o \leqq 0$. The background action is zero for $\tilde{U}_o = 0$. In this work, we do not check for the special case $\tilde{S}^{\mathrm{cs}}=0$. Simply, the action has a negative value for the dS space. It is related to the fact that the Euclidean action for Einstein gravity is not bounded from below, and this is known as the conformal factor problem in Euclidean quantum gravity \cite{ghp98}. It was argued in \cite{dalo} that the conformal divergence due to the unboundedness of the action might get
cancelled with a similar term of opposite sign caused by the measure
of the path integral. However, the difference between the action of
the solution and that of the background remains positive-valued.

\subsection{Oscillating Fubini instantons }

\begin{figure}[!]
\begin{center}
\includegraphics[width=5.in]{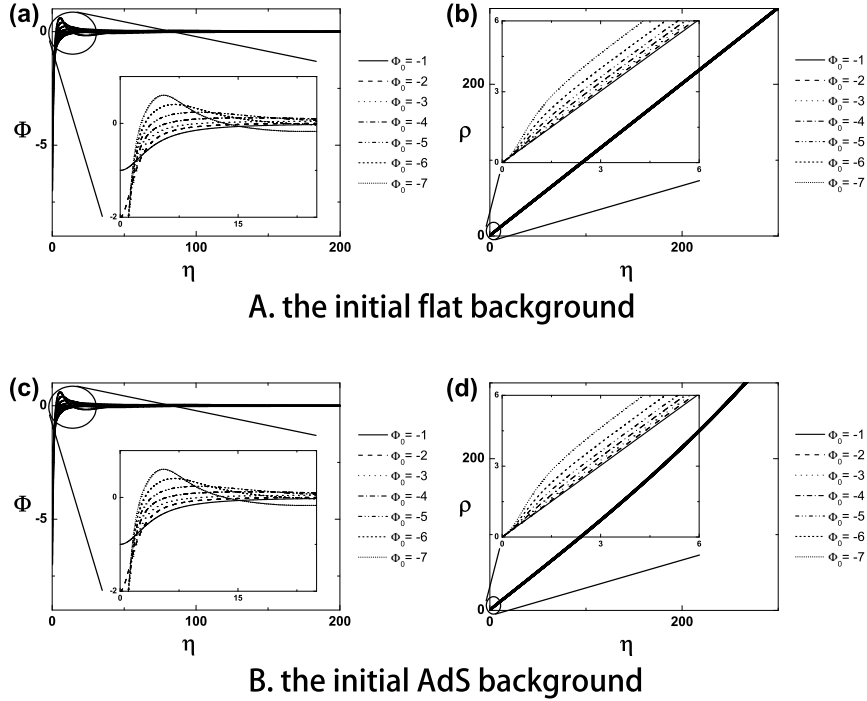}
\end{center}
\caption{\footnotesize{The numerical solutions
representing oscillatory solutions. }} \label{fig:fig06}
\end{figure}

The oscillating instanton and the bounce solutions with an $O(4)$
symmetry between the dS-dS vacuum states was first studied in Ref.\
\cite{hw000}, in which the authors found the solutions in a fixed
background geometry and showed how does the maximum allowed number
$n_{\mathrm{max}}$ depend on the parameters of the theory, where $n$
denotes the crossing number of the potential barrier by the
oscillating solutions. The oscillating bounce solutions in the
presence of gravity was also studied in Ref.\ \cite{nemode}, in
which the authors analyzed the negative modes and the fluctuations
around the oscillating solutions. The instanton was interpreted as the thermal tunneling \cite{brwein}. The oscillating instanton
solutions under a symmetric double-well potential in the curved
space with an arbitrary vacuum energy was also investigated in
detail in \cite{lllo2}, where a numerical solution is possible as
long as the local maximum value of the potential remains positive.
The solutions have a thick wall and can be interpreted as a
mechanism for the nucleation of the thick wall for topological
inflation \cite{avil04}. Similarly, the process for the tunneling
without a barrier in curved space, was studied in Ref.\
\cite{klee08, ljps}. The existence of numerical solutions was shown
in Ref.\ \cite{lllo2}, in which the case representing the tunneling
from flat to AdS space shows an oscillating behavior. The solution
oscillates around $\Phi=0$ in the inverted potential and the
oscillating behavior die away unlike the case under a harmonic
potential.

In the present paper, the oscillation means that the field in the
solutions oscillates around the minimum of the inverted potential
and die away asymptotically to the minimum $\Phi=0$ for the case
with $m^2=0$. Thus the resulting geometry of the initial state has
wrinkles due to the variation of the volume energy density and the
instanton simultaneously. The behavior of the solutions representing
the resulting geometry with wrinkles is quite different from those
in Ref.\ \cite{lllo2}.

Figure \ref{fig:fig06} shows the numerical solutions representing an
oscillatory behavior in (A) the initial flat background and (B) the initial AdS background. We take $\tilde{\kappa} =0.30$, $\tilde{U}_o = 0$ (for the flat case), and $\tilde{U}_o = -0.0001$ (for the AdS case), respectively. Figures \ref{fig:fig06}(a) and (c) illustrate the numerical solutions of $\tilde{\Phi}$. The lower right box in those figure shows the magnification of a small region representing the behavior of the solution around $\tilde{\Phi}=0$. The peak corresponds to the first turning point of the particle similar to a classical mechanics problem in the presence of an inverted potential. For the case with $\tilde{\Phi}_o = -7$ the first turning point reaches furthermost point away from $\tilde{\Phi}=0$ among all the other cases, as one can easily see from the figure. The curves oscillate around $\tilde{\Phi}=0$ and eventually stop at $\tilde{\Phi}=0$ in the flat and AdS space. We take the initial point as an arbitrary $\tilde{\Phi}_o$, which means that the number of oscillations for each solution can be different. However, there is the tendency that the number of oscillations is decreased as the value of $\tilde{\Phi}_o$ is decreased in the given $\tilde{\kappa}$. Figures \ref{fig:fig06}(b) and (d) illustrate the numerical solutions for $\tilde{\rho}$. The upper left box in those figure shows the magnification of an initial small region representing behavior of the curves which move below with the decrease in $\tilde{\Phi}_o$.
\begin{table}
\begin{center}
\renewcommand{\arraystretch}{1.5}
\begin{tabular}{ccccccc}
  \noalign{\hrule height0.8pt}
  $\tilde{\Phi}_{0}$ & $S^{\mathrm{cs}}$ (AdS) & $S^{\mathrm{bg}}$ (AdS) & B (AdS) & B (flat) \\
  \hline
  $-1$ & $9.03462\times10^5$ & $9.01815\times10^5$ & $1.64726\times10^3$ & $1.08561\times10^2$ \\
  $-2$ & $9.05239\times10^5$ & $9.01815\times10^5$ & $3.42429\times10^3$ & $1.58782\times10^2$ \\
  $-3$ & $9.07584\times10^5$ & $9.01815\times10^5$ & $5.76952\times10^3$ & $2.77028\times10^2$ \\
  $-4$ & $9.10850\times10^5$ & $9.01815\times10^5$ & $9.03545\times10^3$ & $3.62471\times10^2$ \\
  $-5$ & $9.16368\times10^5$ & $9.01815\times10^5$ & $1.45537\times10^4$ & $7.67645\times10^2$ \\
  $-6$ & $9.26167\times10^5$ & $9.01815\times10^5$ & $2.43521\times10^4$ & $1.37407\times10^3$ \\
  $-7$ & $9.47248\times10^5$ & $9.01815\times10^5$ & $4.54330\times10^4$ & $3.70685\times10^3$ \\
  \noalign{\hrule height0.8pt}
\end{tabular}
\end{center}
\caption{\footnotesize{The dimensionless variables and color of plot
used and the actions obtained in Fig.\ \ref{fig:fig06}.}}
\label{table3}
\end{table}

Table \ref{table3} shows the initial values of $\tilde{\Phi}$, the actions for the AdS, and flat background which are obtained from Fig.\ \ref{fig:fig06}. In the flat case, the background action is zero as $\tilde{U}_o=0$ and therefore $\tilde{S}^{\mathrm{cs}}$ is equal to $\tilde{B}$. In the present case, we cut all the data at a certain point which is $\tilde{\eta}=200$.

\begin{figure}
\begin{center}
\includegraphics[width=5.in]{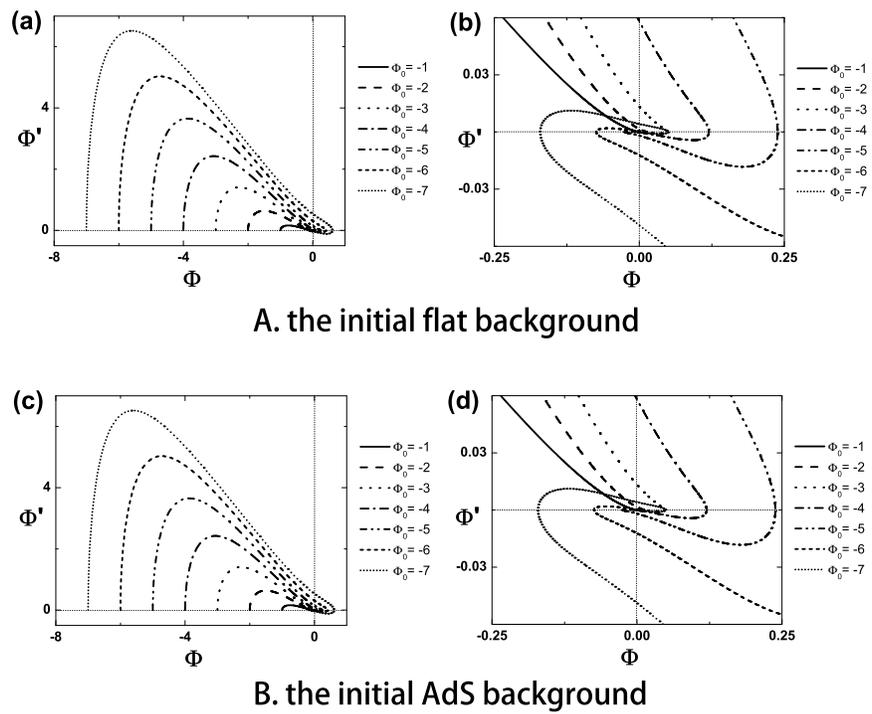}
\end{center}
\caption{\footnotesize{The behavior of the solutions
represented in the phase diagram. }} \label{fig:fig07}
\end{figure}

We now analyze the behavior of the solutions using a phase diagram
method. After plugging the value of $\frac{\rho'}{\rho}$ from Eq.\
(\ref{eqcon}) into Eq.\ (\ref{erho}) and using
$\Phi''=\Phi'\frac{d\Phi'}{d\Phi}$, the equation becomes
\begin{equation}
\frac{d\Phi'}{d\Phi} = - \frac{3\sqrt{\frac{1}{\rho^2} +
\frac{\kappa}{3}(\frac{1}{2}
\Phi'^2+\frac{\lambda}{4}\Phi^4-U_o)}\Phi' +\lambda\Phi^3}{\Phi'}\,.
\label{ephadi}
\end{equation}
First, we consider the situation where the kinetic energy is small
compared to the potential energy such that $|U| \gg \Phi'^2$, $U_o
\ll 1$ and the term $1/\rho^2$ is smaller than other terms. In other
words, the last term is the most dominant among other terms in the
numerator. Then the equation reduces to the form
\begin{equation}
\Phi' \simeq \sqrt{\frac{\lambda}{2}(\Phi^4_o-\Phi^4)} \,. \label{fereduce}
\end{equation}
The above relation shows that the first stage of the curve has got
such kind of form. Second, we consider the situation where
$d\Phi'/d\Phi=0$, i.e. with vanishing acceleration and then we
impose all the above mentioned conditions. It will then describe the
special points in the phase diagram. Then the equation reduces to
the form
\begin{equation}
\Phi' \simeq - 2\sqrt{\frac{\lambda}{3\kappa}}\Phi\,. \label{ereduce}
\end{equation}
Third, we consider the situation where $d\Phi'/d\Phi= -c $, i.e. a
negative constant. We impose all the above mentioned conditions
among with $\Phi^2 \gg 2c/\sqrt{3\kappa\lambda}$. Thus, we obtain
the above equation again. This relation implies that the special
points with a vanishing acceleration and some of the region with a
negative constant acceleration in the phase diagram have got a
linear function type behavior in the phase diagram as shown in
Figs.\ \ref{fig:fig07}(a) and (c).

Figure \ref{fig:fig07} illustrates the behavior of the solutions in
the $\tilde{\Phi}$-$\tilde{\Phi}'$ plane. Each trajectory represents the behavior of the solution in the phase diagram. The trajectories begin with zero velocity as $\tilde{\Phi}'=0$ shown in Figs.\ \ref{fig:fig07}(a) and (c). The velocity increases rapidly to the maximum and then decreases linearly up to the turning point. Figures \ref{fig:fig07}(b) and (d) show the magnification of the small region representing the behavior of the solution around $\tilde{\Phi}'=0$ and $\tilde{\Phi}=0$.

Basically, the Fubini solution has an asymptotic condition to be
satisfied. We expect that there exist an oscillating solutions
although the dS background has got a finite size in the Euclidean
signature. However, if we consider an analytic continuation not of
the angle parameter $\chi$ but of the Euclidean evolution parameter
$\eta=it$, then the meaning becomes clearer. When there is an `even'
symmetry for the oscillating instantons, we can see the half-way
point $\eta_{0}$ as $\dot{\rho}(\eta_{0})=\dot{\Phi}(\eta_{0})=0$.
Then, we can paste the Lorentzian manifold $t=0$ at the
$\eta=\eta_{0}$ surface. This is possible only for the case
$\dot{\rho}(\eta_{0})=\dot{\Phi}(\eta_{0})=0$, because of the
Cauchy-Riemann theorem of complex analysis; otherwise, the
Lorentzian manifold should be complex valued functions (for
exceptional cases, we might be able to consider complex valued
instantons, the so-called fuzzy instantons \cite{nbmtu, hsyl}). In
this procedure, an event shows a spontaneous creation of the
universe from `nothing' \cite{eterinf}, in which nothing means a
state without the concept of classical spacetime \cite{vil141}. We
already know that there is such a solution when the scalar field is
exactly on top of the local maximum. However, now we observe a
creation from nothing with highly non-trivial field dynamics. This
is worthwhile to be highlighted and we postpone further analysis for
the future work.

\subsection{Fubini instantons with $Z_2$ symmetry}

We now shift our attention to the new type of solutions in the initial background as the dS space, i.e. $U_o > 0$. The Euclidean dS space has a compact geometry. Thus the solutions can have $Z_2$ symmetry. We consider the boundary conditions in Eq.\ (\ref{ourbc-3}). To obtain the solutions with $Z_2$ symmetry, we need to impose additional conditions. For the background geometry, $\rho'=0$ at $\eta=\frac{\eta_{max}}{2}$. On the other hand, for the scalar field, we impose $\Phi=0$ at $\eta=\frac{\eta_{max}}{2}$ for the solutions with odd number of crossings of the potential well and $\Phi'=0$ at $\eta=\frac{\eta_{max}}{2}$ for the solutions with even number of crossings. The solutions with odd number of crossing have the opposite state of the value $\Phi$ at $\eta=0$ and $\eta=\eta_{max}$, i.e. $\Phi|_{\eta=\eta_{max}}=-\Phi_o$. The solutions with even number of crossing have the same state of the value $\Phi$ at $\eta=0$ and $\eta=\eta_{max}$, i.e. $\Phi|_{\eta=\eta_{max}}=\Phi_o$. We stress that the boundary conditions in Eq.\ (\ref{ourbc-3}) gives rise to
completely new type of solutions of Fubini instanton.

\begin{figure}[!]
\begin{center}
\includegraphics[width=6.in]{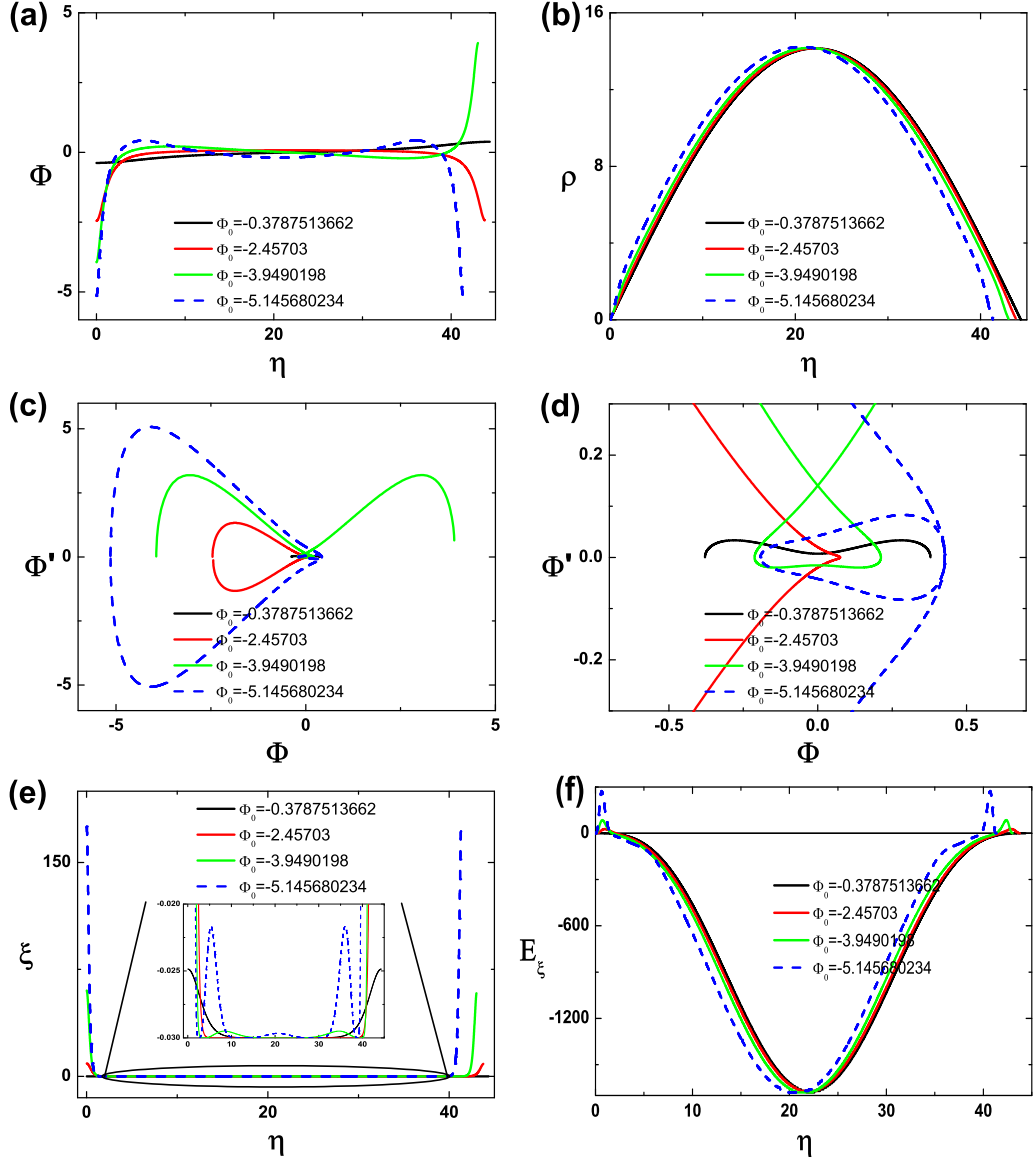}
\end{center}
\caption{\footnotesize{(color online). The numerical solutions
of the Fubini instanton with $Z_2$ symmetry.}}
\label{fig:fig08}
\end{figure}

Figure \ref{fig:fig08} shows the numerical solutions of the Fubini
instanton with $Z_2$ symmetry. We take $\tilde{\kappa} =0.50$ and $\tilde{U}_o = 0.03$. Thus the dS region in the $\tilde{\Phi}$-space spans the region $ -0.589 \lesssim \tilde{\Phi} \lesssim 0.589$. We consider four cases with different initial positions of $\tilde{\Phi}$. Figure \ref{fig:fig08}(a) illustrates the numerical solution of $\tilde{\Phi}$. The trajectories with the blue and red color go back to the same position of $\tilde{\Phi}$ in the presence of the inverted potential, i.e. they have even number of crossings. The trajectories with the black and green color go back to the opposite position of $\tilde{\Phi}$, i.e. they have odd number of crossings. Figure \ref{fig:fig08}(b) depicts the numerical solution of $\tilde{\rho}$. Figures \ref{fig:fig08}(c) and (d) illustrate the behavior of the solutions in the $\tilde{\Phi}$-$\tilde{\Phi}'$ plane. Each trajectory represents the behavior of the solution in the phase diagrams. The blue and red lines indicate that the interior part of two instantons has got the same state as $\tilde{\Phi}$, whereas the green and black lines indicate that the interior part has got the opposite state of $\tilde{\Phi}$. Figure \ref{fig:fig08}(d) illustrates the magnification of a small region representing the behavior of the solution around $\tilde{\Phi}'=0$ and $\tilde{\Phi}=0$. Figure \ref{fig:fig08}(e) illustrates the volume energy density, where the density has got a form $\tilde{\xi}=-\tilde{U}$. The box shows the magnification of a small region representing behavior of curves. The densities in each of the case have got positive values near the initial starting point $\tilde{\Phi}_o$ far away from the point $\tilde{\Phi}=0$, because the densities have the form $\tilde{\xi}=-\tilde{U}$ and $\tilde{U}_o > 0$. The solutions oscillate in the dS region as found the present work. The density always negative values for the case of the black line. Figure \ref{fig:fig08}(f) illustrates the Euclidean energy
$\tilde{E}_{\tilde{\xi}}=2\pi^2\tilde{\rho}^3\tilde{\xi}$ for each slice of constant $\tilde{\eta}$ values.
The negative energy parts in each of the case signifies a rolling
state in the dS region. Table \ref{table4} shows the initial value
of $\tilde{\Phi}$, colors of the plot used, and the actions obtained from
Fig.\ \ref{fig:fig08}.

\begin{table}
\begin{center}
\renewcommand{\arraystretch}{1.5}
\begin{tabular}{ccccc}
  \noalign{\hrule height0.8pt}
  $\tilde{\Phi}_{0}$ & Color of plot & $S^{\mathrm{cs}}$ & $S^{\mathrm{bg}}$ & B \\
  \hline
  $-0.37875$ & Black & $-3.15525\times10^4$ & $-3.15827\times10^4$ & $30.2$ \\
  $-2.45703$ & Red & $-3.15176\times10^4$ & $-3.15827\times10^4$ & $65.2$ \\
  $-3.94902$ & Green & $-3.14005\times10^4$ & $-3.15827\times10^4$ & $182.2$ \\
  $-5.14568$ & Blue & $-3.10727\times10^4$ & $-3.15827\times10^4$ & $510.0$ \\
  \noalign{\hrule height0.8pt}
\end{tabular}
\end{center}
\caption{\footnotesize{The dimensionless variables and the color of
plot used, and the actions obtained in Fig.\ \ref{fig:fig08}.}}
\label{table4}
\end{table}

\section{Causal structures \label{sec4}}

In this section, we briefly outline the causal structure of the
solutions in the Lorentzian signature. Due to the pressure
difference, the nucleated AdS region will expand over the background
and hence the boundary of the nucleated AdS region will be
time-like.

\begin{figure}
\begin{center}
\includegraphics[width=6.in]{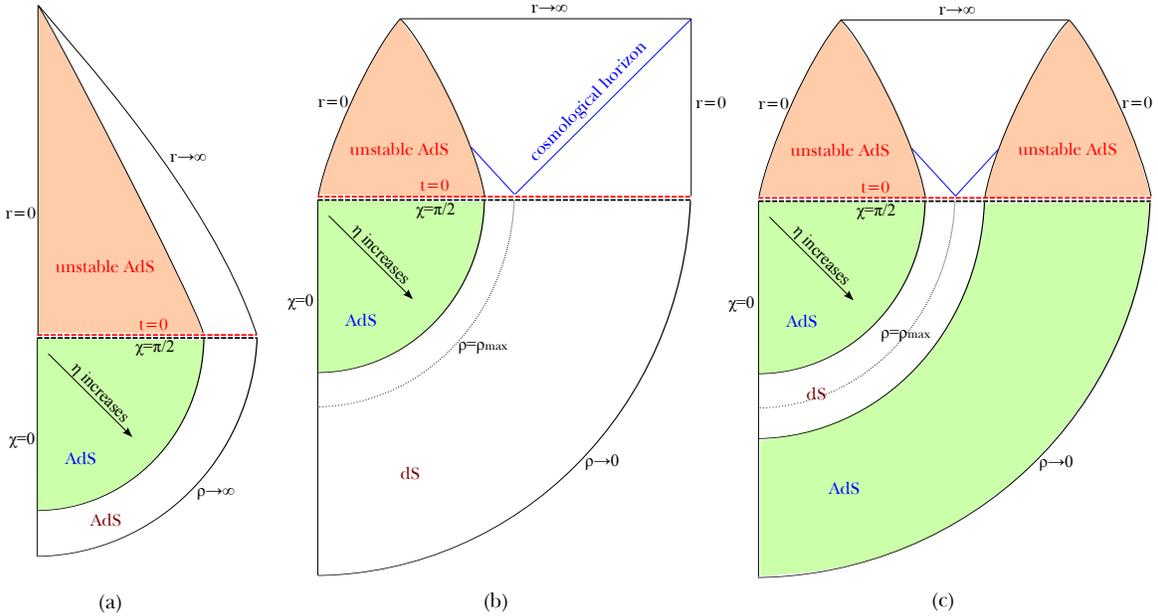}
\end{center}
\caption{\footnotesize{(color online). The schematic diagrams
representing the causal structure of the Fubini instantons and the
related solutions. (a) Tunneling in an AdS background. (b)
oscillating instanton solution in a dS background. (c) Pair creation
by oscillating instantons in dS background.}} \label{fig:fig09}
\end{figure}

Figure \ref{fig:fig09} shows the schematic diagrams representing the
causal structures of the Fubini instantons and the related
solutions. The $\chi=\pi/2$ surface can be analytically continued to
the surface $t=0$ in the Lorentzian signature. The lower vacuum
region in the instanton (green colored region) will be unstable
during the Lorentzian time evolution (orange colored region). Due to
the instability of the Fubini type potential, the whole causal
structure may depend on the shape of the potential or the vacuum
structure i.e. whether the left or the right side of the potential
has true vacua or not. Therefore, the followings are meaningful only
as reasonable estimations for general behavior and these may be
different for special examples.

Figure \ref{fig:fig09}(a) illustrates the instanton solution in an
AdS background. It will form time-like $r=0$ and $r=\infty$
boundaries in the Lorentzian signature. However, the AdS region may be unstable to form a kind of singularity. Figure
\ref{fig:fig09}(b) illustrates the instanton solution in the dS
background. The dS region has a cosmological horizon and will this
form a future infinity. The AdS region (orange colored region) will
expand over the dS region due to the pressure difference. Figure
\ref{fig:fig09}(c) is the pair creation by the oscillating instanton
solutions. Therefore, in the instanton part, the dS region around
the $\rho=\rho_{\mathrm{max}}$ is surrounded by the AdS (green
colored region) part. In the Lorentzian signature, we can interpret
these two AdS parts as being nucleated in a dS background. In Fig.\
\ref{fig:fig09}(c), we infer that, there still remains a dS region
and a future infinity.

The pair creation of the instantons in this work is
quite different from the ordinary quantum process of pair creation
of particles. We take the initial background as the dS space, i.e.
$U_o > 0$. The Euclidean dS space has a compact geometry. Thus, the
geometry has two poles. If one object is created on the north pole
and the other on the south pole, we can interpret that process as
the pair creation of objects. As an example of the process, the
two-crossing solution between the sides of the potential barrier in
the double-well potential was considered as a type of a
double-bounce solution or an anti-double-bounce solution
\cite{rlpr}, in which the authors interpreted the double-bounce
solution as the spontaneous pair-creation of true vacuum bubbles,
one at each pole in the dS space. We adopt a similar interpretation
for our solutions with $Z_2$ symmetry.

\section{Summary and Discussions \label{sec6}}

In this paper we have studied Fubini instantons of a
self-gravitating scalar field representing the tunneling without a
barrier. There are two kinds of Euclidean solutions representing the
tunneling without any barrier. One of them is the tunneling from the
local maximum of the potential to the vacuum state. The other one is
the tunneling from the maximum to any arbitrary state. The latter
corresponds to the Fubini instanton solution. We have shown that
there exist several new kinds of Fubini instanton solutions of a
self-gravitating scalar field found as numerical solutions, which
are possible only if gravity is taken into account. We also computed
the action difference $B$, in each case, between the Euclidean
action corresponding to a classical solution $S^{\mathrm{cs}}$ and
the background action $S^{\mathrm{bg}}$ for the rate of decay.

In Sec.\ 2, we reviewed the Fubini instanton in the absence of
gravity from the viewpoint of a tunneling problem. We have presented
a numerical solution including the Euclidean energy density for
example. We analyzed the structure of the solution in a theory with
the potential having only a quartic self-interaction term. These
solutions can be considered as a ball consisting of only a thick
wall except for the one point at the center of the solution with a
lower arbitrary state than the outer vacuum state unlike a vacuum
bubble which consists of an inner part with a lower vacuum state and
a wall.

In Sec.\ 3, we have studied the instanton solutions in curved
space. We performed careful numerical study to solve the coupled
equations for the gravity and the scalar field simultaneously. We
have shown that there exist numerical solutions without any
oscillation in the initial AdS space for the potential with only the
quartic term. We have also shown that there exist numerical
solutions for the potential with both a quartic and a quadratic term
irrespective of the value of the cosmological constant. For this
particular case, there is no solution with an $O(4)$ symmetry when
gravity is switched off. In order to estimate the decay rate of the
background state, we calculated the action difference between the
action of the solution and that of the background obtained using
numerical means.

We have obtained oscillating Fubini instantons as new
types of solutions. We have shown that there exist oscillating
numerical solutions for the potential with only the quartic term in the flat and AdS space, except for the solution without oscillation in the initial AdS space with the specific value of a cosmological constant and the parameters. We have analyzed the behavior of the solutions using the phase diagram method. The oscillation dies away asymptotically in both the flat and the AdS space.

We have obtained numerical solutions representing the Fubini instanton with $Z_2$ symmetry. We stress that they represent
completely new type of solutions of Fubini instanton. These solutions can be interpreted as the pair creation with each one having the same state and with each one having the opposite state, respectively. The solutions can lead to more interesting interpretation as follows: any arbitrary state can tunnel into another arbitrary state with an $O(4)$-symmetry in the curved spacetime, although no vacuum state exists as the instanton solution. The solutions are possible as long as the maximum of the potential remains positive.

The subject on the pair creation of bubbles was first considered in Ref.\ \cite{game}. The numerical solution representing the pair of solutions is in Fig.\ 2 in Ref.\ \cite{rlpr}, which can be interpreted as the pair creation of the bubbles, one at each pole in the dS space. However, there is a different interpretation on the solutions \cite{brwein, aliwein}, in which the authors studied a decay channel of de Sitter vacua. The solutions with $O(3)$ symmetry can be understood as describing tunneling in a finite horizon volume at finite temperature. The solutions maybe correspond to thermal production of a bubble in their interpretation. In this stage, the comparative analysis between the O(4)-symmetric solution and O(3)-symmetric solution with respect to the pair creation is needed to be studied more. We leave this for future work.

In Sec.\ 4, we have analyzed the schematic diagrams representing the causal structures of the Fubini instantons and the related solutions
in the Lorentzian signature. For the special case representing the solution with $Z_2$ symmetry, the dS region around the $\rho=\rho_{\mathrm{max}}$ is surrounded by an AdS part.

We now mention on the negative mode problem. It was known that the bounce solution has one negative mode in the spectrum of small perturbations about the solution \cite{ccol, nemo0}. The bounce solution with one negative mode corresponds to the tunneling process in the lowest WKB approximation. In Ref.\ \cite{nemo0}, Coleman argued that the Euclidean solution with only one negative mode is related to the tunneling process in the flat Minkowski spacetime. However, there is no rigorous proof on extension of Coleman's argument to the curved space claiming the physical irrelevance of the solutions with additional negative modes. For example, the time-translation invariance, or zero modes, is one of crucial elements to prove the uniqueness of the negative mode in his argument. However, the existence of zero modes is not guaranteed in curved spacetime. Another point is that the Euclidean time interval is at most of $O(H^{-1})$ in de Sitter space. Hence, only a finite number of the bounces can be placed far apart from each other. Therefore, the dilute gas approximation may become invalid easily, which leads to the breakdown of the WKB approximation \cite{nemo4}.
There appears diverse situations on the negative modes when the gravity is taken into account \cite{nemo1, nemo2, nemo4, nemo5}. Although, the bounce solution with one negative mode in curved space dominates the tunneling process, the solutions with additional negative modes may also contribute to the tunneling process. There exist some works including the physical interpretation on the oscillating solutions with more than one negative mode. One can naturally interpret the system in de Sitter background as a thermal system. The authors in Refs.\ \cite{hw000, brwein} interpreted that the existence of additional negative modes represents the solutions as unstable intermediate thermal configuration. They seem to observe the clue to support this idea on the other point of view. It is known that the $N$ times oscillating solutions have $N$ negative modes \cite{hw000, brwein, nemo3}. The even numbers of negative modes of the form $4N$ and $4N+2$ do not have imaginary part of the energy, while the odd numbers of negative modes of the form $4N+3$ have the imaginary part of the energy with the wrong sign. However, $4N+1$ negative modes may have a meaning for a tunneling process even if the solution may not be related to the lowest WKB approximation. Recently the analysis on the negative modes of oscillating instantons has been investigated \cite{nemo3}. The oscillating instantons as homogeneous tunneling channels have been also studied \cite{bwdy9}. In  conclusion, we believe many Euclidean solutions in curved space with zero and negative modes may have physical significance and deserves further investigation.

In summary, we illustrate the following finding in our new
contribution regarding this issue:

\begin{enumerate}

\item In the absence of gravity, a $-\phi^{4}$-type potential has infinitely many instanton solutions whereas a $-\phi^{4}+\phi^{2}$-type potential has no instanton solution. However, \textit{the inclusion of the gravity changes all the situation abruptly}: for the former case, the solution space get reduced to a finite space and for the latter case, there exists solutions.

\item We also confirm that $-\phi^{4}$-type potentials have oscillating instanton solutions as well as the solutions with $Z_2$ symmetry.

\end{enumerate}

Therefore, the Fubini instanton is one of the few examples that
shows the effect of gravity bringing drastic changes to the
tunneling process. There can be more applications of the oscillating
instantons and we confirm that the Fubini-type potentials also
contribute largely towards these processes. We postpone any possible
application of such oscillating solutions including the phase space of solutions for our future work \cite{llry}.

\section{Acknowledgements}

We would like to thank Andrei Linde for kind historic comments on
the inflationary multiverse scenario and Fubini instanton. We would
like to thank Erick J. Weinberg, George Lavrelashvili, Hongsu Kim, Yunseok Seo, and Dong-il Hwang for helpful discussions and comments, and thank Chaitali Roychowdhury for a careful English revision of the manuscript. We would like to thank Manu B.~Paranjape and Richard MacKenzie for their hospitality during our visit to Universit\'{e} de Montr\'{e}al. This work was supported by the Korea Science and Engineering Foundation (KOSEF) grant funded by the Korea government(MEST) through the Center for Quantum Spacetime(CQUeST) of Sogang University with grant number R11 - 2005 - 021. WL was upported by Basic Science Research Program through the National Research Foundation of Korea(NRF) funded by the Ministry of Education, Science and Technology(2012R1A1A2043908). DY is supported by the JSPS Grant-in-Aid for Scientific Research (A) No. 21244033. We appreciate APCTP for its hospitality during completion of this work.

\end{document}